# A System for Smart-Home Control of Appliances Based on Timer and Speech Interaction

S. M. Anamul Haque[1], S. M. Kamruzzaman[2] and Md. Ashraful Islam[1]

Department of Computer Science & Engineering,
[1]International Islamic University Chittagong.
[2]Manarat International University.
E-mail: anam_ecs@yahoo.com

**ABSTRACT**

**The main objective of this work is to design and construct a microcomputer based system: to control electric appliances such as light, fan, heater, washing machine, motor, TV, etc. The paper discusses two major approaches to control home appliances. The first involves controlling home appliances using timer option. The second approach is to control home appliances using voice command. Moreover, it is also possible to control appliances using Graphical User Interface. The parallel port is used to transfer data from computer to the particular device to be controlled. An interface box is designed to connect the high power loads to the parallel port. This system will play an important role for the elderly and physically disable people to control their home appliances in intuitive and flexible way. We have developed a system, which is able to control eight electric appliances properly in these three modes.**

## 1. INTRODUCTION

Personal computers are increasingly becoming the platform of choice to design and implement control algorithms because it is simple to write, modify and update software programs that implement a control algorithm. In this project, we use the personal computer to control the electric appliances. For example turning high power AC loads such as lights, fans, heater etc. ON or OFF. To successfully integrate the digital controller with the analog plant, an interface device is used within the PC that can perform the necessary tasks [1]. In this document, we present the design of the interface box. The interface box can be controlled by the computer by connecting to the parallel port and written a program in any preferable language. Here we write a program in Visual basic programming language. The program will demonstrate the basic idea of how to control devices and monitor events. By the program we can have the computer to turn electric devices ON/OFF while away from home by using timing option. More over the people who are physically disabled in home and work place are able to control the home appliances using voice commands.

The paper first describes the general architecture of the home-appliances control system. The details hardware of the interface box, which is the most important part for interfacing high power loads with the computer, is described in section 3. In section 4, we describe the software development phase. Finally, we represent the Graphical User Interface (GUI) of the system software and summarize our work in the remaining sections.

## 2. ARCHITECTURE OF THE PROPOSED SYSTEM

The aim of this project is to be controlled home appliances by using a personal computer. The overall system architecture is shown in fig.1.

**Fig. 1**: Architecture of the proposed system

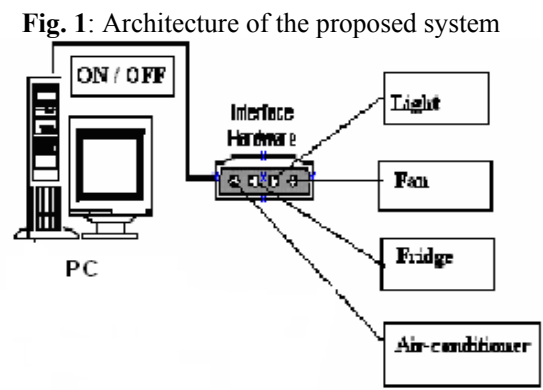

The system consists of two units: a Control unit and an interface unit. The control unit is based on the use of standard personal computer with Graphical User Interface (GUI) software to control the electrical appliances. The Interface unit is for interfacing the high power loads with the control unit.

## 3. HARDWARE DESIGN OF THE INTERFACE BOX

In this section, we present the design of the interface box that is used to connect high power loads to the computer. Details about the components that comprise the interface box are also shown here. We have shown how to interface a relay and an AC load

### 3.1 External Panel

The upper panel of the interface box consists of sockets, which are easily accessible to the electric appliances for proper and safe operation. A snapshot of the external panel of the interface box is shown in fig.2.

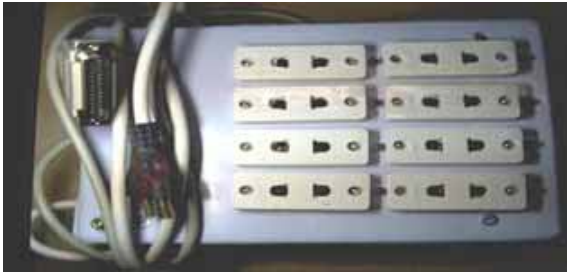

**Fig. 2:** External panel of an interface box

The Front Panel consists of an AC power supply Cable, Power ON switch, Printer cable, Sockets, and LEDs. AC power supply cable connects the interface box to the main AC supply. Power ON switch supplies 220V (50Hz) mains' supply to the interface box when turned on and stop supplies when turned off. Printer cable connects the interface box to the PC's parallel port. Electric appliances are generally connected to Socket. The LEDs indicate the status of the Electric appliances (i.e. ON or OFF).

### 3.2 Internal Modules

The Internal circuitry of the interface box can be divided into three main categories: Relay driver Circuit, Relays and +6V DC power supply.

### 3.2.1 Relay Driver Circuit

The relay [2] driver circuit is necessary to drive the 6V DC relay. The circuit diagram for each unit in the relay driver with relay and socket is shown in fig.3. The relay driver circuit is necessary for each socket.

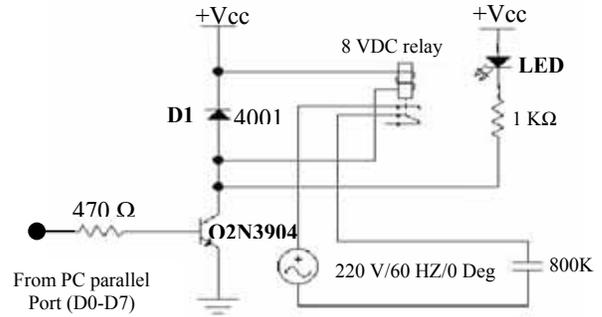

**Fig. 3:** Circuit diagram for the relay driver with relay and socket.

### 3.2.2 Relays

A total of eight relays are used for the eight output sockets. Eight relays are wired in such a way that the normally closed pin of the relay connects to the Power Resistors and the Normally Open pin connects to a socket.

### 3.2.3 6V DC Power Supply

The DC power supply is needed to provide +6V DC source to the Relay driver circuit.

### 3.3 Computer Interfacing

PC's parallel port is a 25 pins D-shaped female connector in the back of the computer [3], [4]. This is an inexpensive and yet powerful platform for implementing projects dealing with the control of real world peripherals. The parallel port is made up of three different sections. These are data lines, control lines and status lines. There are 8 data lines, and they are the primary means of getting information out of the port. The control lines are another 4 outputs. The status lines are a standard parallel port's inputs. There are 5 of them. The fig.4 shows all the pins of the parallel port. The pins 18,19,20,21,22,23,24 and 25 are all ground pins.

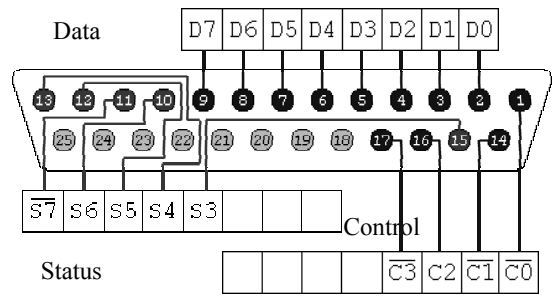

**Fig. 4:** Pin outlines of the parallel port

**Table 1:** Parallel port address

| Address | Notes |
|---|---|
| 3BCh-3BFh | Does not support ECP address |
| 378h-37Fh | Usually address for LPT1 |
| 278h-27Fh | Usually address for LPT2 |

LPT1 is normally assigned base address 378h while LPT2 is assigned 278h. The address may change from machine to machine. The addresses are given in Table-1. The 8 data lines output pins are adequate for the project.

Those data pins are TTL level output pins. This means that they put out ideally 0V when they are in low logical level (0) and +5V when they are in logical level (1). We use the logical level (1) to ON the electrical appliances and logic level (0) to OFF. Data pin 0 is used to control the device 1; pin 2 is used to control the device 2 and so on.

### 4. SOFTWARE DEVELOPMENT

We have implemented our software using Visual Basic 6.0 as programming language and Microsoft voice engine tools for speech recognition purpose. The develop software can be run under Windows 98/ME/NT/2000/2000 Server/XP.

**4.1 Device Driver**

Visual Basic can not directly access the hardware on a system [5], [6]. All hardware requests must go through Windows. In order to control the port directly, we used something external to our program. Writing programs to talk with parallel port was pretty easy in old **DOS** version and in **WIN95/98** too. We use **win95IO.dll** file and the functions **VbOut()** and **VbInp()** for outputting and inputting data to and from the parallel port respectively. But entering to the new era of NT clone operating systems like **WINNT/WIN2000/WINXP** all this simplicity goes away. Windows NT assigns some privileges and restriction to different types of program running on it. It classifies all the programs into two categories: *user mode* and *kernel mode*. Device drivers are capable of running in kernel mode. We have write a device driver to handle the "Interface Box" by using the **Inpout32.dll** file and **Inp32()** and **Out32()** function for **WINNT/2000/XP** operating system**.**

**4.2 Speech Recognition**

Speech recognition, or speech-to-text, involves capturing and digitizing the sound waves, converting them to basic language units or phonemes, constructing words from phonemes, and contextually analyzing the words to ensure correct spelling for words that sound alike (such as write and right) [7]-[9]. The fig.5 illustrates this high-level description of the process.

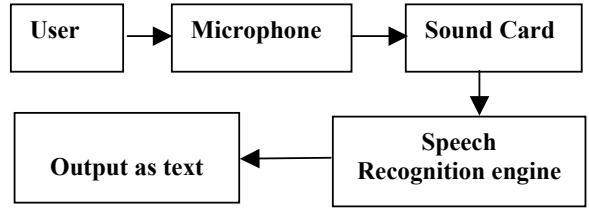

**Fig. 5:** Block diagram of speech recognition process flow

A number of English recognition engines are available now. Microsoft has also been developing voice recognition for a few years. We have used Microsoft Speech Recognition Engine (version 4.0) for the recognition of English speech in our system.

### 5. GRAPHICAL INTERFACE

The graphical interfaces of the developed software are smart enough to guide a person to extract full functionality of the utility program without any prior experience. The Main window provides three options to control the electric appliances as shown in fig.6.

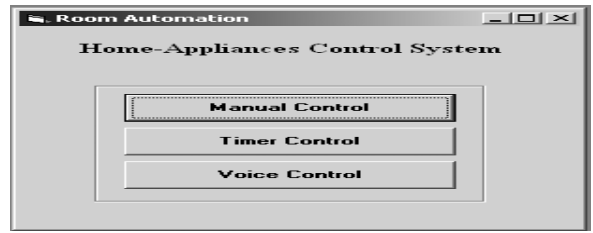

**Fig. 6:** Main window

**5.1 Manual Control Window**

This window contains ON and OFF push buttons for each of the electric appliances to turn ON or OFF respectively, as shown in fig.7.

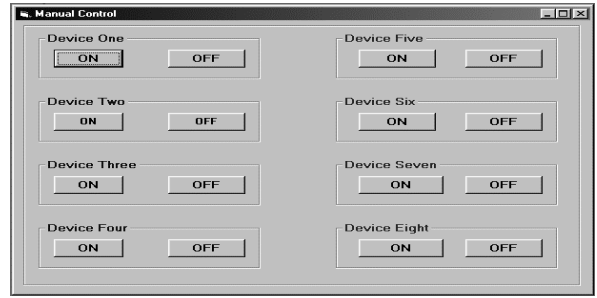

**Fig. 7:** Manual control window

## 5.2 Timer Control Window

This window provides the facilities to set date and time and status of the device operation (ON or OFF) to turn ON or OFF the devices according to the given date and time as shown in fig.8.

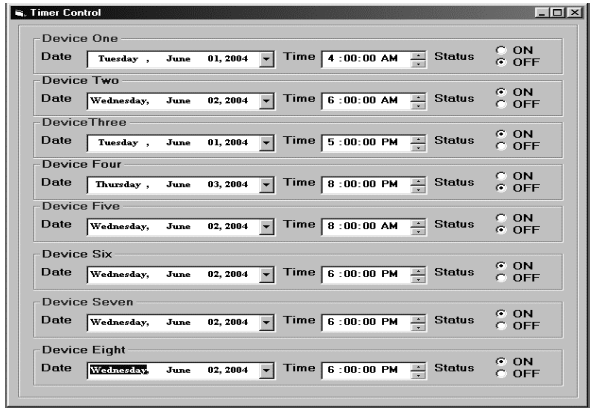

**Fig. 8:** Timer control window

## 5.3 Voice Control

The voice control window shown in fig.9 will provide the opportunity to control home appliances with the help of Voice command.

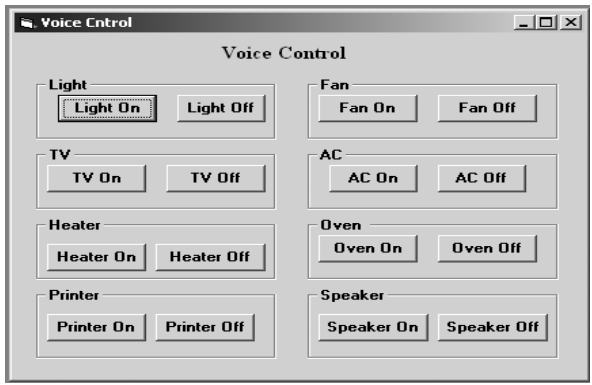

**Fig. 9:** Voice control window

The user needs only to utter the words with in the Command button to turn ON/OFF the specific device. Suppose, anybody wants to turn on the light, he needs only to utter the word "LightOn".

## 6. CONCLUSION

The developed system is a robust combination of a number of diverse technologies, to construct a speech activated portal to home appliances to assists users who are disable and in home and workplaces. This system can control only eight electric appliances and can be further control 256 electric devices in industry or large organization through a single PC by upgrading the interface box and a little modification in the program.